\documentstyle[11pt,cite]{article}

\def\al           {\alpha}
\def\be           {\beta}

\def\de           {\delta}
\def\ep           {\epsilon}

\def\ga           {\gamma}

\def\la           {\lambda}

\def\Ga           {\Gamma}

\def\Pi           {\Pi}

\def\non          {\nonumber}

\def\ha           {\mbox{$\frac{1}{2}$}}

\def\d        {\mbox{d}}
\def\spa          {\ \ \ }
\def\mand         {\spa\mbox{and}\spa}

\def\Tr           {\mbox{\rm Tr}\,}
\def\STr          {\mbox{\rm STr}\,}

\def\cd           {{\cdot}}
\def\ran          {\rangle}
\def\lan          {\langle}

\def\fsH	{H\!\!\!\!/\,}

\scrollmode

\begin{document}

\pagenumbering{arabic}
\setcounter{page}{0}

\hfill hep-th/9905195
$\mbox{}$

\vspace{3cm}
\begin{center}
{\Large{\bf Ramond-Ramond Couplings on Brane-Antibrane Systems}}

\vspace{1cm}
\renewcommand{\thefootnote}{\fnsymbol{footnote}}
Conall Kennedy\footnote{E-mail: conall@maths.tcd.ie} and
Andy Wilkins\footnote{E-mail: andyw@maths.tcd.ie}
\setcounter{footnote}{0}

{\em Department of Mathematics \\
Trinity College \\
Dublin 2 \\
Ireland}

(May 1999)

\vspace{5mm}

\begin{minipage}{14cm}
Couplings between a closed string RR field and open strings are
calculated in a system of coincident branes and antibranes of type II
theory.  The result can be written cleanly using the curvature of the
superconnection. 
\end{minipage}
\end{center}

\vfill 
\noindent

\newpage

\section{Introduction}

Unstable D-brane systems can decay and produce branes of lower dimension.
The most simple decay is that of a D$p$-brane annihilating with a
D$p$-antibrane to yield a D$(p-2)$-brane.  In this 
scenario the lightest open-string mode stretching between the
brane-antibrane pair is tachyonic and it can condense into a
configuration with non-trivial winding.  Then, in order to
have finite energy, the gauge field living on the branes must have
non-zero first Chern class which implies the existence of a non-zero
$(p-2)$-brane charge.  The happy state of affairs is that the negative
energy density of the condensed tachyon field exactly cancels with the
positive energy density of the brane-antibrane pair asymptotically,
leaving a D$(p-2)$-brane with finite tension~\cite{sen3}.

One can also start with non-supersymmetric D$p$-branes in type II
theory ($p$ is odd for IIA and even for IIB).  Here too there is a
tachyon in the spectrum of the open string connecting the brane to
itself.  It can condense into a kink and the unstable
non-supersymmetric brane will decay to yield a stable D$(p-1)$
brane. This type of brane production was studied in detail
in~\cite{hor1} where it was argued that all supersymmetric D-branes in
IIA can be constructed as bound states of a number of D9-branes.  
Further examples of such decay processes are studied
in~\cite{ber2,sen7,sen2,sen8,hor1,sen1,fra1} and can be understood
within the context of K-theory~\cite{wit1,ber1}.   The unstable and
non-BPS branes have also been used in testing various
duality conjectures~\cite{sen5,sen6,sen3,sen7}, indeed it was in this
context that they made their first major appearance.

The configurations that are studied in this note are a straightforward
generalisation of the scenario described in the first paragraph to $N$
D$p$-branes coincident with $N$ D$p$-antibranes.  Without the
antibranes the coupling of the U($N$) 
massless world-volume vectors (denoted by $A$) to the closed-string RR
fields (denoted by $C$) is given by the `Wess-Zumino'
action~\cite{polchinski1995a,li1996a}
\begin{equation}
S = \int_{\Sigma_{(p+1)}} C \wedge \Tr e^{F} \ ,
\label{eqn.wz}
\end{equation}
where $\Sigma_{(p+1)}$ is the world volume, $\Tr$ is over the
Chan-Paton factors and $F$ is the field strength for $A$.  Numerical
factors such as $\al'$ have been suppressed, as have the contributions from
the antisymmetric 2-form and the A-roof
genus~\cite{bershadskyET1996a,greenET1997a,cheungET1998a}.   When 
the antibranes are included, the light degrees of freedom are two
U($N$) gauge fields --- one living on the branes ($A^{+}$) and the
other on the antibranes ($A^{-}$) --- and a tachyon ($T$) and
antitachyon ($\bar{T}$) living in the $(N,\bar{N})$ and $(\bar{N},N)$
representations respectively.   We 
would like to know the generalisation of the Wess-Zumino action.
To this end we perform a tree-level string calculation of the 
effective action to low orders in the tachyon field.

We find a non-zero contact interaction of the form
\begin{equation}
\la\int_{\Sigma_{(p+1)}} C_{(p-1)}\wedge \d \,\Tr
\left( T\wedge \overline{DT} \right) \ ,
\label{eqn.fir}
\end{equation}
where the covariant derivatives are
\begin{displaymath}
DT = dT + A^{+}T - TA^{-} \mand
\overline{DT} = d\bar{T} - \bar{T}A^{+} + A^{-}\bar{T} \ .
\non\end{displaymath}
In this note we will not calculate the overall normalisation of such
terms (except to say their coefficients are not zero) since they are
unimportant for our purposes.   We also consider the 
brane-antibrane pairs to have indistinguishable world volumes.\footnote{Further
comments on this point are made in~\cite{pes1} in which the tachyon
potential was shown to assume a Mexican hat shape for weak fields.}
The result Eq.~(\ref{eqn.fir}) means the total charge of the
$(p-2)$-brane is measured by $\int \Tr \left(F^{+} - F^{-} + \la\d (T
\overline{DT}) \right)$ which contains the first Chern class of the
gauge configuration on the brane, the antibrane, and the winding
number of the tachyon configuration,   However, this latter term 
does not add unwanted charge to the $(p-2)$-brane because the
covariant derivative, $DT$, must vanish at infinity in
order for the solitonic configuration to have finite energy.

The coupling to the RR $(p-1)$-form given by Eqs.~(\ref{eqn.wz})
and~(\ref{eqn.fir}) can be rewritten to read
\begin{equation}
\int_{\Sigma_{(p+1)}} C_{(p-1)} \wedge 
\Tr \left ( F^{+} - F^{-} - \ha  \left\{
F^{+},T\bar{T}\right\}
+\ha \left\{ F^{-}, \bar{T}T \right\} +
  DT\wedge \overline{DT} 
\right)  \ ,
\label{eqn.ful}
\end{equation}
Interestingly, this can be written in a more compact
form by employing the superconnection of noncommutative
geometry~\cite{quil,berl,roep}
\begin{displaymath}
{\cal A} = \left(
\begin{array}{cc}
d + A^{+} & T \\ \bar{T} & d + A^{-} 
\end{array}
\right) \ ,
\non\end{displaymath}
which transforms under the U($N$)$\times$U($N$) symmetry as
\begin{displaymath}
{\cal A} \rightarrow  {\cal GAG}^{-1}
\spa \mbox{where}\spa {\cal G} = \left
( \begin{array}{cc}g&0\\0&\bar{g}\end{array} 
\right)
\ .
\non\end{displaymath}
The curvature of this is
\begin{displaymath}
{\cal F} = \left(
\begin{array}{cc}
F^{+} - T\bar{T} & DT \\
\overline{DT} & F^{-} - \bar{T}T  
\end{array}
\right) \ ,
\non\end{displaymath}
and a ``supertrace'' is defined by
\begin{displaymath}
\STr \left( \begin{array}{cc} a&b\\c&d \end{array} \right)
= \Tr a - \Tr d \ .
\non\end{displaymath}
Then the terms in Eq.~(\ref{eqn.ful}) are just those in an expansion
of
\begin{equation}
S = \int_{\Sigma_{(p+1)}}C \wedge \STr e^{\cal F} \ .
\label{eqn.pbarp}
\end{equation}
We propose that this generalises the usual
Wess-Zumino action Eq.~(\ref{eqn.wz}).

Before moving on to the string calculation, let us show that our
proposal gives the correct charges for all decay products.  This can
be seen by using the `transgression formula'~\cite[p.~47]{berl} which
reads
\begin{displaymath}
\STr \exp{\cal F}_{1} - \STr \exp{\cal F}_{0} = \d  \int_{0}^{1}dt\, \STr
 \frac{\d {\cal A}_{t}}{\d t} \exp{\cal F}_{t}
\ ,
\non\end{displaymath}
where ${\cal A}_{t}$ is a superconnection depending on a continuous
parameter $t$, and ${\cal F}_{t}$ is its curvature.  Let us choose
\begin{displaymath}
{\cal A}_{t} = \left( \begin{array}{cc} d + A^{+}&tT\\ t\bar{T}& d + A^{-}
\end{array}\right) \ .
\non\end{displaymath}
When the transgression formula is integrated over the $2k$ directions
perpendicular to the $(p-2k)$-brane's world volume we can set
$DT=0=\overline{DT}$ on the RHS.  The RHS then integrates to zero because
it contains only {\em odd} dimensional forms.   This implies 
\begin{displaymath}
\int_{2k}\STr\exp{\cal F} = \int_{2k}\Tr\exp F^{+} - \int_{2k}\Tr \exp
F^{-}  \ ,
\non\end{displaymath}
as required.

\section{The calculations}

The process of interest is one where a RR boson annihilates onto the
brane-antibrane world volume to create some open strings.
By string duality this is just an open string graph with insertions on
the boundary (the brane's world volume) and one RR insertion in the
interior.  Similar calculations 
have been performed in~\cite{gub1,gar1,hash2,hash1,gar3,gar2} and we
will follow their notations and conventions:

Map the disc to the upper-half plane so the boundary of the disc
becomes the real axis.  On this axis the worldsheet bosons ($X^{\mu}$)
and fermions ($\psi^{\mu}$) obey the following boundary conditions
\begin{displaymath}
\mbox{Neumann: }\left\{
\begin{array}{rcl}
X^{a}(z) & = & \tilde{X}^{a}(\bar{z})
\\
\psi^{a}(z) & = & \tilde{\psi}^{a}(\bar{z})
\end{array}
\right. \mand \mbox{Dirichlet: }\left\{
\begin{array}{rcl}
X^{i}(z) & = & - \tilde{X}^{i}(\bar{z}) 
\\
\psi^{i}(z) & = & -\tilde{\psi}^{i}(\bar{z})
\end{array}
\right. \ ,
\non\end{displaymath}
while the ghosts ($c$) and superghosts ($\phi$) obey trivial boundary
conditions.   Indices $\mu = 0,\ldots,9$, while an index $a$ lives on
the world volume ($a = 0,\ldots ,p$) and $i$ fills the transverse
space ($i = (p+1),\ldots,9$).  Because of these boundary conditions the
correlators mix; 
\begin{equation}
\begin{array}{rrclcrcl}
\mbox{Neumann: } & \lan X^{a}(z)X^{b}(w) \ran & = &
-\eta^{ab}\log(z-w) & \mand & \lan X^{a}(z)\tilde{X}^{b}(\bar{w}) \ran & =
& -\eta^{ab} \log(z-\bar{w})
\ , \\
\mbox{Dirichlet: } & \lan X^{i}(z)X^{j}(w) \ran & = &
-\eta^{ij}\log(z-w) &\mand & \lan
X^{i}(z)\tilde{X}^{j}(\bar{w}) \ran & = & +\eta^{ij} \log(z-\bar{w})
\ .
\end{array}
\label{eqn.mixed.correlators}
\end{equation}
and similarly for the fermions (we use the conventions $\al' =2$).
Now use the ``doubling trick'' in 
which the fields $X(z)$ and $\psi(z)$ are extended to the lower-half
plane by defining
\begin{displaymath}
X(z) = \left\{
\begin{array}{lcl}
\tilde{X}(z) && \mbox{Neumann} \\
-\tilde{X}(z) && \mbox{Dirichlet}
\end{array}
\right. \spa (z\in\,\mbox{LHP}) \ .
\non\end{displaymath}
Then if we think of $\bar{w}$ being in the LHP the correlation function
of this extended holomorphic field 
\begin{displaymath}
\lan X^{\mu}(z)X^{\nu}(w)\ran = -\eta^{\mu\nu}\log(z-w) \ ,
\non\end{displaymath}
correctly reproduces all of Eq.~(\ref{eqn.mixed.correlators}).  Thus,
when considering scattering off $p$-branes, the rule is to replace
\begin{displaymath}
\tilde{X}^{\mu}(\bar{z}) \rightarrow D^{\mu}_{\nu}X^{\nu}(\bar{z}) \ ,
\spa
\tilde{\psi}^{\mu}(\bar{z}) \rightarrow
D^{\mu}_{\nu}\psi^{\nu}(\bar{z}) \ ,
\spa
\tilde{\phi}(\bar{z}) \rightarrow \phi(\bar{z}) \mand
\tilde{c}(\bar{z}) \rightarrow c(\bar{z}) \ ,
\non\end{displaymath}
where
\begin{displaymath}
D = \left( \begin{array}{cc}
1_{p+1} & 0 \\
0 & -1_{9-p}
\end{array}
\right) \ ,
\non\end{displaymath}
and then use the usual correlators
\begin{eqnarray}
\lan X^{\mu}(z)X^{\nu}(w)\ran & = & -\eta^{\mu\nu}\log(z-w) \ , \non \\
\lan \psi^{\mu}(z)\psi^{\nu}(w) \ran & = & -\eta^{\mu\nu}(z-w)^{-1} \ ,
\non \\
\lan c(z)c(w) \ran & = & (z-w) \ , \non \\
\lan\phi(z)\phi(w)\ran & = & -\log(z-w) \ .
\non\end{eqnarray}

The vertex operators for the tachyon are
\begin{displaymath}
V_{T}^{(0)}(x) = k\cd\psi e^{ik\cd X}(x)
\mand
V_{T}^{(-1)}(x) = e^{-\phi} e^{ik\cd X}(x) \ ,
\non\end{displaymath}
where the superscripts label the superghost number.  The momentum $k$
is constrained to lie in the world volume; $k^{\mu} = (k^{a},0)$ with
$k^{2} = 1/4$ in our conventions ($\al' =2$).  In the coincident
brane-antibrane system the vertex operators for the tachyon and the
antitachyon look the same --- in order to distinguish them a
Chan-Paton factor must be understood.  After doubling, the vertex
operators become~\cite{kos1} 
\begin{eqnarray}
V_{T}^{(0)}(z) & = & k\cd\psi e^{2ik\cd X}(z)
\ , \non\\
V_{T}^{(-1)}(z) & = & e^{-\phi} e^{2ik\cd X}(z) \ .
\non\end{eqnarray}
The RR vertex operators are
\begin{eqnarray}
V_{RR}^{(-1)}(w,\bar{w}) & = & (P_{-}\fsH_{(m)})^{\al\be}\,
:e^{-\phi/2}S_{\al}e^{ip\cd X}(w):\,
:e^{-\tilde{\phi}/2}\tilde{S}_{\be}e^{ip\cd \tilde{X}}(\bar{w}):
\non\end{eqnarray}
with the projector $P_{-} = \ha (1-\ga^{11})$ assuring that we are using
the correct chirality and
\begin{displaymath}
\fsH_{(m)} = \frac{1}{m!}H_{\mu_{1}\ldots\mu_{m}}\ga^{\mu_{1}}\ldots
\ga^{\mu_{m}}
\ ,
\non\end{displaymath}
where $m=2,4$ for type IIA and $m=1,3,5$ for type IIB.  The spinorial
indices are raised with the charge conjugation matrix, eg
$(P_{-}\fsH_{(m)})^{\al\be} =
C^{\al\de}(P_{-}\fsH_{(m)})_{\de}{}^{\be}$ (further conventions and
notations for spinors can be found in appendix~B of~\cite{gar1}).  The
RR bosons are massless so $p^{2}=0$.  The spin fields can also be
extended to the entire complex plane.  In calculations we replace 
\begin{displaymath}
\tilde{S}_{\al}(\bar{w}) \rightarrow M_{\al}{}^{\be}{S}_{\be}(\bar{w})
\ ,
\non\end{displaymath}
where
\begin{displaymath}
M = \frac{1}{p!}\ga^{a_{0}}\ga^{a_{1}}\ldots \ga^{a_{p}}
\ep_{a_{0}\ldots a_{p}}  \ . 
\non\end{displaymath}
Finally, a couple of correlators containing two spin fields are
\begin{eqnarray}
\lan S_{\al}(w)S_{\be} \ran & = & w^{-5/4}C^{-1}_{\al\be}
\ , \non \\
\lan \psi^{\mu}(z) S_{\al}(w)S_{\be} \ran & = &
(z-w)^{-1/2}z^{-1/2}w^{-3/4}\ga^{\mu}_{\al\be} 
\ .
\non\end{eqnarray}
where $S_{\be}=S_{\be}(0)$ and $C_{\al\be}$ is the charge conjugation
matrix. 

\subsection{The two-point amplitude}

The amplitude containing one RR field and one tachyon (or antitachyon) is
\begin{eqnarray}
{\cal A}^{T,RR} & = & \int \frac{dzdwd\bar{w}}
{V_{\mbox{\scriptsize{CKG}}}}  \lan V_{T}^{(-1)}(z)
V_{RR}^{(-1)}(w,\bar{w})\ran
\ , \non \\
& = &  \int \frac{dzdwd\bar{w}}
{V_{\mbox{\scriptsize{CKG}}}}  \lan 
e^{-\phi(z)}e^{2ik\cd X(z)}
e^{-\phi(w)/2} S_{\al}(w)  e^{ip\cd X(w)}
(P_{-}\fsH_{(m)}M)^{\al\be}
\non \\
&& \spa\spa\spa\spa\spa\spa\spa\spa
\times e^{-\phi(\bar{w})/2} S_{\be}(\bar{w}) e^{ip\cd D \cd X(\bar{w})}
 \ran \ .
\non\end{eqnarray}
We have chosen the vertex operators according to the rule that the
total superghost number must be~$-2$.
Including the ghost contribution,
$
\lan c(z)c(w)c(\bar{w})\ran = (z-w)(z-\bar{w})(w-\bar{w})
$,
and using the kinematic constraint, $k^{a}+p^{a}=0$, the volume of the
conformal Killing group can be canceled by fixing the three insertion
points.  The amplitude then reads
\begin{displaymath}
{\cal A}^{T,RR} = \Tr \left( P_{-}\fsH_{(m)}M \right) \ .
\non\end{displaymath}
Repeated use of
$\{\ga^{\mu},\ga^{\nu}\} = 2\eta^{\mu\nu}$ yields
\begin{displaymath}
\Tr \left( \ga^{[\mu_{1}}\ldots \ga^{\mu_{m}]} 
 \ga_{[a_{0}}\ldots \ga_{a_{p}]} \right)
= 32(p+1)!\de_{m,p+1}
 \de^{[\mu_{m}}_{[a_{0}}\de^{\mu_{m-1}}_{a_{1}} \ldots
 \de^{\mu_{1}]}_{a_{p}]} 
\ ,
\non\end{displaymath}
implying that the amplitude vanishes since there is no $H_{(p+1)}$ in
the type II string.  On the other 
hand, in the case of a non-BPS brane (as studied recently
in~\cite{bill}) this amplitude implies the effective action contains
the term $\int C\wedge \d T$.

\subsection{The three-point amplitude}

The three-point amplitude between one RR field, and two tachyonic
particles is
\begin{eqnarray}
{\cal A}^{T,T,RR} & = & \int \frac{dzdz'dwd\bar{w}}
{V_{\mbox{\scriptsize{CKG}}}}  \lan V_{T}^{(0)}(z)
V_{T}^{(-1)}(z')
V_{RR}^{(-1)}(w,\bar{w})\ran
\ , \non \\
& = &  \int \frac{dzdz'dwd\bar{w}}
{V_{\mbox{\scriptsize{CKG}}}}  \lan 
k\cd\psi(z) e^{2ik\cd X(z)}e^{-\phi(z')}e^{2ik'\cd X(z')}
e^{-\phi(w)/2} S_{\al}(w)  e^{ip\cd X(w)}
(P_{-}\fsH_{(m)}M)^{\al\be}
\non \\
&& \spa\spa\spa\spa\spa\spa\spa\spa
\times e^{-\phi(\bar{w})/2} S_{\be}(\bar{w}) e^{ip\cd D \cd X(\bar{w})}
 \ran \ .
\non\end{eqnarray}
It is convenient to fix the points $(z,z',w,\bar{w}) = (x,-x,i,-i)$ to
cancel the volume $V_{CKG}$ by inserting the ghost contribution
\begin{displaymath}
\lan c(z')c(w)c(\bar{w}) \ran = (z'-w)(z'-\bar{w})(w-\bar{w}) \ .
\non\end{displaymath} 
Introduce the Mandelstam variable $t = -(k+k')^2$.  Then using various
kinematic constraints such as 
\begin{displaymath}
k^{a} + k'{}^{a} + p^{a} = 0 \mand
p\cd D\cd p = -2t = -2p\cd (k+k') \ ,
\non\end{displaymath}
the amplitude reduces to
\begin{eqnarray}
{\cal A}^{T,T,RR} & = & \int_{-\infty}^{\infty} \d x
\left( \frac{(1+x^{2})^{2}}{16 x^{2}}\right)^{\ha + t} \frac{1}{1+x^{2}} \Tr
(P_{-}\fsH_{(m)}M\ga^{a})k_{a} \ , \non \\
& = & \pi \frac{\Ga[-2t]}{\Ga[\ha-t]^2}
 \Tr (P_{-}\fsH_{(m)}M\ga^{a})k_{a} \ .
\label{eqn.tt.rr}
\end{eqnarray}
Next the trace must be evaluated;
\begin{displaymath}
\Tr \left( \ga^{\mu_{1}}\ldots \ga^{\mu_{m}} 
\ga^{a_{0}}\ldots \ga^{a_{p}} \ga^{a}
 \right)H_{\mu_{1}\ldots\mu_{m}}\ep_{a_{0}\ldots a_{p}}
=  32(p+1)!\de_{m,p}(-1)^{p(p+1)/2}H_{a_{0}\ldots a_{p-1}}
 \ep^{a_{1}\ldots a_{p-1}a} \ .
\non\end{displaymath}
The trace containing the factor of $\ga^{11}$ ensures the following
results also hold for $p>3$ with $H_{(m)} \equiv \ast H_{(10-m)}$ for
$m\geq 5$. 

The prefactor of Eq.~(\ref{eqn.tt.rr}) has the interesting property that
at non-negative integer values of $t$ there is a pole corresponding to
an open-string resonance with mass-squared $m^{2} = t = -p_{a}^{2}$.
On the other hand, at positive half-integer values of $t$ it vanishes,
implying that strings with half-integer mass-squared do not propagate
in this channel. In~\cite{hash2} it was shown that these are also
properties of the amplitude for one NS-NS string to decay into two
massless open strings stuck to a D-brane.

The low-energy effective Lagrangian of massless and tachyonic
particles will contain the terms
\begin{equation}
{\cal L}_{\mbox{\scriptsize{eff}}} \sim H_{(p)}\wedge A + {\cal
 L}_{\mbox{\scriptsize{YM}}} 
 + DT\cd\overline{DT}  \ ,
\label{eqn.rr.a.t}
\end{equation}
as well as the terms we are looking for.  Here ${\cal
L}_{\mbox{\scriptsize{YM}}}$ is the Yang-Mills Lagrangian and in
keeping  with the spirit of the rest of this note all constants have
been omitted. At low energies ($-t = p_{a}^{2} \sim 0$) the
prefactor of Eq.~(\ref{eqn.tt.rr}) may be expanded
\begin{displaymath}
\pi\frac{\Ga[-2t]}{\Ga[\ha-t]^2}
 =  \frac{-1}{2t} + 2\log 2 + O(t)
\ .
\non\end{displaymath}
The first term corresponds to the RR particle decaying into a massless
open string (via the first term in Eq.~(\ref{eqn.rr.a.t})) which
propagates (resulting in the pole) and decays into two
tachyons (via the third term in Eq.~(\ref{eqn.rr.a.t})).  Because
the second term is non-zero, the effective action contains a coupling
between the RR field, the tachyon and the antitachyon.

In summary, in this subsection we have shown that the effective action
contains the term 
\begin{displaymath}
S_{\mbox{\scriptsize{eff}}} = \int_{\Sigma_{(p+1)}} H_{(p)}\wedge
\Tr \bar{T}\d T \ . 
\non\end{displaymath}
which is Eq.~(\ref{eqn.fir}) after integrating by parts and
covariantising.  As an aside, if there is no antitachyon (in the case
of the non-supersymmetric brane) the integration by parts gives zero.

%

\section{Summary}

In coincident brane-antibrane systems we have shown, by calculating
tree-level string amplitudes, that in addition to the usual Wess-Zumino
terms the world-volume effective action contains
\begin{displaymath}
\int_{\Sigma_{(p+1)}}C_{(p-1)}\wedge \d\  \Tr (T\overline{DT})
\ ,
\non\end{displaymath}
to $O(T\bar{T})$.  After tachyon condensation the correct charges for
decay products are obtained.   We propose that the full result
(to all orders in the tachyon) can be written in terms of the
curvature of the superconnection: 
\begin{displaymath}
\int_{\Sigma_{(p+1)}}C \wedge \STr \exp{\cal F}  \ . 
\non\end{displaymath}

\section*{Acknowledgments}

We thank Siddhartha Sen and Jim McCarthy for commenting on various
drafts.  This work was supported financially by the Enterprise Ireland
grant F01121.

\end{document}